\documentclass[preprint]{aastex}

\newcommand{\myemail}{cmorgan@usna.edu}
\newcommand{\avgm}{\langle M \rangle}

\shorttitle{Quasar Accretion Disk Sizes}
\shortauthors{Morgan et al.}

\begin{document}

\title{The Quasar Accretion Disk Size -- Black Hole Mass Relation\footnote{Based on 
observations obtained with the Small and Moderate Aperture Research Telescope 
System (SMARTS) 1.3m, which is operated by the SMARTS Consortium,
the Apache Point Observatory 3.5-meter telescope, which is 
owned and operated by the Astrophysical Research Consortium, the WIYN Observatory
which is owned and operated by the University of Wisconsin, Indiana University, 
Yale University and the National Optical Astronomy Observatories (NOAO), the 
6.5m Magellan Baade telescope, which is a collaboration between the observatories 
of the Carnegie Institution of Washington (OCIW), University of Arizona, 
Harvard University, University of Michigan, and Massachusetts Institute 
of Technology, and observations made with the NASA/ESA Hubble Space Telescope
for program HST-GO-9744 of the Space Telescope Science Institute,
which is operated by the Association of Universities for Research
in Astronomy, Inc., under NASA contract NAS 5-26555.
}}

\author{Christopher W. Morgan}
\affil{Department of Physics, United States Naval Academy, 572C Holloway Road,
Annapolis, MD 21402}
\email{\myemail}
 
\author{C.S. Kochanek and Nicholas D. Morgan}
\affil{Department of Astronomy, The Ohio State University, 140 West 18th Avenue, Columbus, OH 43210
-1173}
\email {ckochanek@astronomy.ohio-state.edu, nmorgan@astronomy.ohio-state.edu}

\and

\author{Emilio E. Falco}
\affil{Harvard-Smithsonian Center for Astrophysics, 60 Garden Street, Cambridge, MA, 02138}
\email{efalco@cfa.harvard.edu}

\clearpage

\begin{abstract}
We use the microlensing variability observed for eleven 
gravitationally lensed quasars to show that the accretion disk size at 
a rest-frame wavelength of 2500\AA~is related to the black hole mass by 
$\log(R_{2500}/$cm$)=(15.78\pm0.12) + (0.80\pm0.17)\log(M_{BH}/10^9$M$_{\sun})$.  
This scaling is consistent 
with the expectation from thin disk theory ($R \propto M_{BH}^{2/3}$), but when 
interpreted in terms of the standard thin disk model ($T \propto R^{-3/4}$), it implies that black 
holes radiate with very low efficiency, $\log(\eta)  = -1.77\pm0.29 + \log(L/L_E)$
where $\eta=L/(\dot{M}c^2)$.  Only by making the maximum reasonable shifts in the average inclination,
Eddington factors and black hole masses can we raise the efficiency estimate to be marginally
consistent with typical efficiency estimates ($\eta \approx 10\%$). With one exception, these 
sizes are larger by a factor of $\sim4$ than the size needed to produce the observed 
$0.8\,$\micron~quasar flux by thermal radiation from a thin disk with the same $T \propto R^{-3/4}$ 
temperature profile.  While scattering a significant fraction of the disk emission
on large scales or including a large fraction of contaminating line emission can
reduce the size discrepancy, resolving it also appears to require that accretion disks have
flatter temperature/surface brightness profiles.   
\end{abstract}


\section{Introduction}
\label{sec:introduction}

Despite nearly 40 years of work on accretion disk physics, the simple \citet{Shakura1973} 
thin disk model, its relativistic cousins \citep[e.g.][]{Page1974,Hubeny1997,Hubeny2001,Li2005}
and more sophisticated implementations \citep[e.g.][]{Narayan1997,DeVilliers2003,Blaes2007} 
remain the standard model despite some observational reservations 
\citep[see][]{Francis1991,Koratkar1999,Collin2002}.
Quasar accretion disks cannot be spatially resolved with ordinary telescopes, so we have 
been forced to test accretion physics through time variability \citep[e.g.][]{VandenBerk2004,
Sergeev2005,Cackett2007} and spectral modeling \citep[e.g.][]{Sun1989,Collin2002,Bonning2007}.  
One notable success is the use of reverberation mapping \citep[e.g.][]{Peterson2004} of quasar broad 
line emission to calibrate the relation between emission line widths and black hole 
masses. The line emission, though, comes from scales much larger than the accretion disk, and
attempts to use similar methods on the continuum emission have had limited success,
largely because quasars show little optical variability on the disk 
light-crossing timescale \citep{Collin2002,Sergeev2005}.   

Gravitational telescopes do, however, provide the necessary resolution to study the 
structure of the quasar continuum source.  Each gravitationally lensed quasar image is observed through a 
magnifying screen created by the stars in the lens galaxy.  Sources that are smaller than the Einstein 
radius of the stars, typically $\sim10^{16}$ cm, show time variable fluxes whose amplitude is 
determined by the source size \citep[see the review by][]{Wambsganss2006}. 
Smaller sources have larger variability amplitudes than 
larger sources. In this paper, we exploit the optical microlensing 
variability observed in eleven gravitationally lensed quasar systems to measure the size of their accretion disks,
and we find that disk sizes are strongly correlated with the masses of their central black holes. 
 
In \S\ref{sec:observations} we describe the monitoring data, the lens models we use based
on Hubble Space Telescope ($HST$) images of each system and our microlensing analysis method.
In \S\ref{sec:results} we describe our accretion disk model and our results for the relationship
between disk size and black hole mass. 
While we analyze and discuss the results in terms of a simplified thin disk model, they can be compared to
any other model by comparing our measurement of the half light radius to that expected from the model
of choice because \citet{Mortonson2005} demonstrate that the half-light radius measured from microlensing is 
essentially independent of the assumed disk surface brightness profile.  The surface 
brightness profile is best probed by measuring the dependence of the microlensing amplitude on wavelength 
\citep[see][]{Anguita2008,Poindexter2008,Agol2009,Bate2008,Floyd2009,Mosquera2009}.  
In \S\ref{sec:discussion}, we discuss the results and their implications for thin accretion disk theory.
All calculations in this paper
assume a flat $\Lambda$CDM cosmology with $h=0.7$, $\Omega_{M}=0.3$ and $\Omega_{\Lambda}=0.7$.

\section{Data and Analysis}
\label{sec:observations}
 
We collected new monitoring data in the systems HE~0435--1223, FBQ~0951+2635, 
HE~1104--1805, SDSS~1138+0314, SBS~1520+530 and Q~2237+030 
in the $r$-, $R$- and $V$-bands on the SMARTS 1.3m using 
the ANDICAM optical/infrared camera \citep{Depoy2003}\footnote{http://www.astronomy.ohio-state.edu/ANDICAM/},
the Wisconsin-Yale-Indiana (WIYN) observatory
using the WIYN Tip--Tilt Module (WTTM)\footnote{http://www.wiyn.org/wttm/WTTM\_manual.html},
the 2.4m Hiltner telescope at the 
MDM Observatory using the MDM Eight-K\footnote{http://www.astro.columbia.edu/~arlin/MDM8K/}, 
Echelle and RETROCAM\footnote{http://www.astronomy.ohio-state.edu/MDM/RETROCAM} \citep{Morgan2005}
imagers and the 6.5m Magellan Baade telescope using IMACS \citep{Bigelow1999}.
We supplemented our monitoring data with published quasar light curves  
from \citet{Paraficz2006}, \citet{Schechter1997}, \citet{Wyrzykowski2003},  
\citet{Ofek2003}, OGLE \citep{Wozniak2000a,Wozniak2000b},
\citet{Gaynullina2005}, and \citet{Poindexter2007}.
We measured the flux of each image by comparison to the flux from reference stars 
in the field of each lens. Our analysis of the monitoring data is described
in detail by \citet{Kochaneketal.2006}. In systems with published time delays, we 
offset the light curves by the delays to eliminate the 
intrinsic source variability. In this paper we also make use of previously reported optical 
accretion disk sizes for the lensed quasars
QJ~0158--4325 \citep{Morgan2008}, SDSS~0924+0219 \citep{Morgan2006}, SDSS~1004+4112 \citep{Fohlmeister2008},
PG~1115+080 \citep{Morgan2008} 
and RXJ~1131$-$1231 \citep{Dai2009}. SDSS~0924+0219, SDSS~1138+0314 and Q~2237+030 do not 
have published time delays but are all quadruply lensed quasars with short ($\la 15$~days) estimated
delays.  Since the typical timescale for microlensing is significantly longer than this ($\ga 1$ year),
we ignored the delays in these systems and used their raw lightcurves in our microlensing analysis.
For QJ~0158--4325 we developed an analysis method that allowed us the simultaneously estimate
the time delays and disk sizes including their mutual uncertainties, the details of which are described
in \citet{Morgan2008}. 

All eleven lenses have been observed in the $V$- (F555W), $I$- (F814W) and $H$-bands (F160W) using
the WFPC2, ACS/WFC and NICMOS instruments on {\it HST}.  We fit these images as combinations of
point sources for the quasars and (generally) de Vaucouleurs models for the lenses as
described in \citet{Lehar2000}.  These provided the astrometry used for lens models
and defined a constant mass-to-light ($M/L$) ratio model for the mass distribution in the 
lens models.  We modeled each system using the {\it GRAVLENS} software package \citep{GRAVLENS}. 
For all systems except the cluster lens SDSS 1004+4112,
we generate a series of ten models starting from a constant $M/L$ model
and then adding an NFW \citep{Navarro1996} halo.  The sequence is parametrized by  
$f_{M/L}$, the mass fraction represented by the visible lens galaxy relative to a 
constant $M/L$ model. In general, we start with the constant $M/L$ model, $f_{M/L}=1$, and then
reduce its mass in increments of $\Delta f_{M/L}=0.1$ with the NFW halo's mass rising
to compensate. For the cluster-lensed quasar SDSS 1004+4112, 
we use the fixed mass model from \citet{Fohlmeister2007}, and we
assume a set of 10 evenly spaced stellar mass fractions in the range $0.1 \leq \kappa_*/\kappa \leq 1.0$.
Thus, our results marginalize over any uncertainties in the dark matter halos of
the lenses. For typical lenses, which we expect to be dark matter dominated, microlensing
favors low stellar mass fractions \citep[e.g.][]{Morgan1115,Dai2009,Chartas2009,Pooley2009,Mediavilla2009}.  
We adopt, however, a very conservative philosophy and include a broad spectrum of models
even though a stronger prior favoring dark-matter dominated models is
relatively easy to justify. We do this so that our estimates of the source sizes
depend as little as possible on other model assumptions.  We have nonetheless
tested the sensitivity of the size estimates on lens model dark matter fraction priors, 
and we have found very weak, if any, dependence.

These lens models provide the convergence $\kappa$, shear $\gamma$ and
stellar surface density $\kappa_*$ needed to define the microlensing magnification
patterns. We assume a lens galaxy stellar mass function $dN(M)/dM \propto M^{-1.3}$ with a 
dynamic range of a factor of 50 that approximates the Galactic disk mass function 
of Gould (2000, see also \citealt{Poindexter2010a}).   We also know from previous theoretical studies
that the choice of the mass function will have little effect on
our conclusions given the other sources of uncertainty \citep[e.g.][]{Paczynski1986,Wyithe2000a}.
For the typical lens we generated 4 magnification 
patterns for each image in each of the 10 lens models. We gave 
the magnification patterns an outer scale of $20 \langle R_E\rangle$, where 
$\langle R_E \rangle$ is the Einstein radius for the mean stellar mass $\langle M\rangle$.
This outer dimension is large enough to fairly sample the magnification pattern, while
the pixel scale of the 4096$^2$ magnification patterns is small enough to resolve the accretion disk.
We determined the properties of the accretion disk by modeling the observed light curves
using the Bayesian Monte Carlo method of \citet{Kochanek2004} \citep[also see][]{Kochanek2007}.  
For a given disk model, we randomly generate light curves, fit them to the observations 
and then use Bayesian methods to compute probability distributions for the disk size
averaged over the lens models, the likely velocities of the observer, lens, source, stars 
and the mean microlens mass $\avgm$.
We used the velocity model from \citet{Kochanek2004}, which used the projected CMB
dipole \citep{Kogut1993} for the observer, a stellar velocity dispersion set by the Einstein radius of the 
lens and peculiar velocity scales for the lens and source of $235/(1+z)$~km s$^{-1}$.
We use a prior on the mean microlens mass of $0.1 {\rm M_\sun} < \langle M \rangle < 1.0{\rm M_\sun}$, 
but the disk size estimates are insensitive to this assumption \citep[see][]{Kochanek2004,Morgan1115}. 

An obstacle to our analysis technique is how to handle differences 
between the observed flux ratios and the flux ratios of the trial lightcurves.  In systems
where we are confident that the intrinsic flux ratios are known 
\citep[e.g. PG 1115+080, for which][measured mid-IR flux ratios]{Chiba2005}, 
we set a prior to favor model lightcurves within 0.1 mag of the 
intrinsic flux ratios.  In most cases, however, the intrinsic flux
ratios are not known because all observed flux ratios can be affected
by chromatic microlensing or dust extinction.  In these cases we set 
a much looser prior on this magnification offset, typically $\sim0.5$~mag.
Loosening the prior on the magnification offset degrades our ability to  
to estimate the dark matter content of the lens galaxy, but it has little effect on accretion disk 
size measurements \citep[see][for a detailed demonstration of this effect in RXJ 1131$-$1231]{Dai2009}.
 
We use black hole mass estimates for the quasars that are based on observed quasar emission 
line widths and the locally calibrated virial relations for black hole masses, for which 
we adopt the combined normalizations of \citet{Onken2004} and \citet{Greene2006}.
For most systems we simply used the black hole mass estimates from \citet{Peng2006} based on
the \ion{C}{4}~($\lambda 1549$\AA), \ion{Mg}{2}~($\lambda 2798$\AA) and H$\beta$~($\lambda 4861$\AA) 
mass-linewidth relations.  For SDSS 1138+0314, we measured the width of the \ion{C}{4}
(1549\AA) line in optical spectra from the Sloan Digital Sky Survey \citep{SDSSDR4} 
and for Q 2237+030 we used the \ion{C}{4} line width measurement from \citet{Yee1991}. We estimated the 
black hole masses for these systems using the virial relation of \citet{Vestergaard2006}. 
For SDSS 1004+4112, we measured the
width of the \ion{Mg}{2}~($\lambda 2798$\AA) emission line in spectra from \citet{Inada2003} and 
\citet{Richards2004} and used the \citet{McLure2002} \ion{Mg}{2} virial relation to estimate its 
black hole mass. These mass estimates
are reliable to approximately 0.3~dex \citep[see][]{McLure2002,Kollmeier2005,Vestergaard2006,Peng2006}. 

\section{Results}
\label{sec:results} 

We model the surface brightness profile of the accretion disk as a power law 
temperature profile, $T \propto R^{-3/4}$, matching the outer regions of a \citet{Shakura1973} thin 
disk model.  We neglect the central depression of the temperature due to the inner edge 
of the disk and corrections from general relativity to avoid extra parameters.  The effect 
of this simplification on our size estimates is small compared to our measurement 
uncertainties provided the disk size we obtain is significantly larger than the 
radius of the inner disk edge. We will compare three disk size estimates in the context
of this simple model.  First, there is our size measurement from the microlensing, 
$R_S$.  This microlensing size should be viewed as a measurement of the half-light radius, 
but we parametrize the results in terms of the simple thin disk model in order to 
facilitate comparisons to the thin disk model. Converted to a half-light radius, $R_{1/2}=2.44R_S$, the measurements
will be nearly model-independent \citep[see][]{Mortonson2005}.  Second, there is the theoretically
expected size as a function of black hole 
mass in the thin disk model (see Eqn.~\ref{eqn:thindisk} below). Third, there is the 
thin-disk size which would yield the observed optical flux assuming thermal radiation
and a $T \propto R^{-3/4}$ temperature profile (see Eqn.~\ref{eqn:fluxsize} below).
We emphasize that while the second and third size estimates
are based upon the same accretion disk model, they differ in that they take different measured
quantities, either $M_{BH}$ or $I$-band flux, as an argument. 
\citet{Collin2002} had previously noticed that the theory size (Eqn.~\ref{eqn:thindisk})
and the flux size (Eqn.~\ref{eqn:fluxsize}) may be discrepant.  \citet{Pooley2007} argued that 
the microlensing sizes and the theory sizes (Eqn.~\ref{eqn:thindisk}) are
discrepant, but they used black hole masses based largely on estimates of
the bolometric luminosity, which in some sense forced a reconciliation
of the flux and theory sizes.

We assume that the disk radiates as a black body, so the surface 
brightness at rest wavelength $\lambda_{rest}$ is
\begin{equation}
f_{\nu} = {2 h_p c \over \lambda_{rest}^3} 
\left[ \exp \left( {R \over R_{\lambda_{rest}}} \right)^{3/4}-1 \right]^{-1}
\label{eqn:disk1}
\end{equation}                                                                                   
where the scale length
\begin{equation}
R_{\lambda_{rest}} = \left[ {45 G \lambda_{rest}^4 M_{BH} \dot{M} \over 16 \pi^6 h_p c^2} \right]^{1/3}
= 9.7 \times 10^{15} \left( {\lambda_{rest} \over {\rm \micron}} \right)^{4/3}
\left( {M_{BH} \over 10^9 {\rm M_{\sun}}} \right)^{2/3}
\left({L \over \eta L_E} \right)^{1/3} {\rm cm}
\label{eqn:thindisk}
\end{equation}
is the radius at which the disk temperature matches the wavelength, 
$k T_{\lambda_{rest}} = h_p c/ \lambda_{rest}$, 
$h_p$ is the Planck constant, $k$ is the Boltzmann constant, $M_{BH}$ is the black hole mass,   
$\dot{M}$ is the mass accretion rate, $L/L_E$ 
is the luminosity in units of the Eddington luminosity, and $\eta=L/(\dot{M}c^2)$ is the accretion efficiency.   
We can also compute the size under the same model assumptions based on the 
magnification-corrected $I$-band quasar fluxes measured in HST observations as
\begin{equation}
R_I = 2.83 \times 10^{15} {1 \over \sqrt{\cos i}} \left( { D_{OS} \over r_H } \right)
\left( { \lambda_{I,obs} \over {\rm \micron} } \right)^{3/2}
10^{-0.2(I-19)} \: h^{-1} \: {\rm cm}
\label{eqn:fluxsize}
\end{equation} 
where $D_{OS}/r_H$ is the angular diameter distance to the quasar in units of the Hubble 
radius, $I$ is the magnification-corrected magnitude and $i$ is the disk inclination angle.

\begin{figure}
\epsscale{1.0}
\plotone{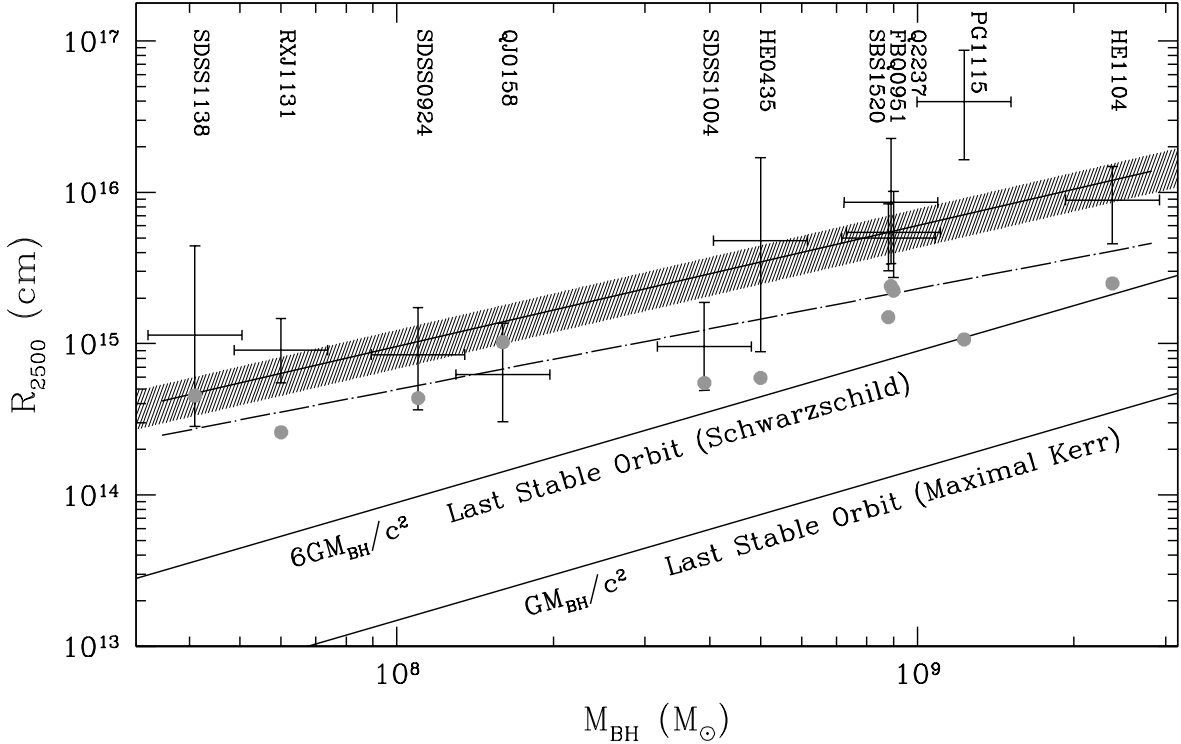}
\caption{Inclination-corrected accretion disk size $R_{2500}$  versus 
black hole mass $M_{BH}$. The solid line through the data shows our best power-law fit to the data 
and the dot-dashed line shows the prediction from thin disk theory ($L/L_E = 1$ and 
$\eta = 0.1$).  The shaded band surrounding the best fit shows the expected
variance due to inclination. 
Disk sizes are corrected to a rest wavelength of  $\lambda_{rest} = 2500$\AA~and the 
black hole masses were estimated using emission line widths.  The filled points 
without error bars are $R_{2500}$ estimates based on the observed, magnification-corrected 
$I$-band fluxes.  They have typical uncertainties of $0.1$-$0.2$~dex.  Solid lines at the
bottom of the plot show the innermost stable circular orbit for a maximally rotating Kerr
black hole and a Schwarzschild black hole, representing a plausible range of radii for the 
inner edge of an accretion disk.  
\label{fig:r2500}}
\end{figure}

\begin{figure}
\epsscale{1.0}
\plotone{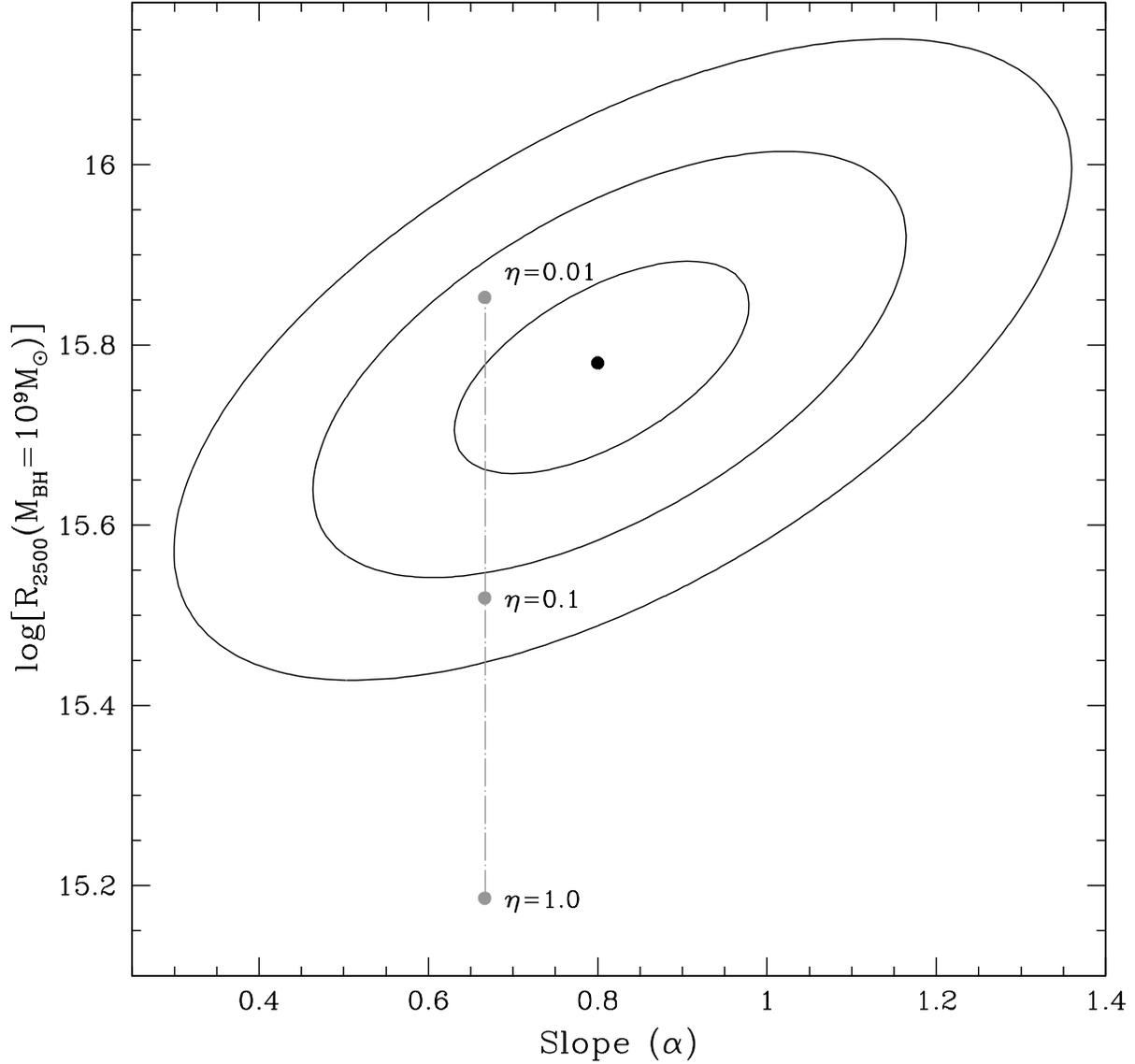}
\caption{Results of the power-law fit to $R_{2500}$ as a function of black 
hole mass.  The contours show the $1-3\sigma$ one-parameter confidence 
intervals for the slope $\alpha$ 
and the normalization $R_{2500}(M_{BH} = 10^9 {\rm M_{\sun}})$ for the 2500\AA~accretion disk size 
corresponding to $M_{BH} = 10^9 {\rm M_{\sun}}$. The best-fit value is indicated with a black point. 
The filled points along the dot-dashed line are theoretical thin disk sizes for quasars radiating at the 
Eddington limit and with efficiencies of $\eta = L / (\dot{M} c^2)  = 0.01$, $0.1$ or $1.0$.
\label{fig:r2500fit}}
\end{figure}

\begin{figure}
\epsscale{1.0}
\plotone{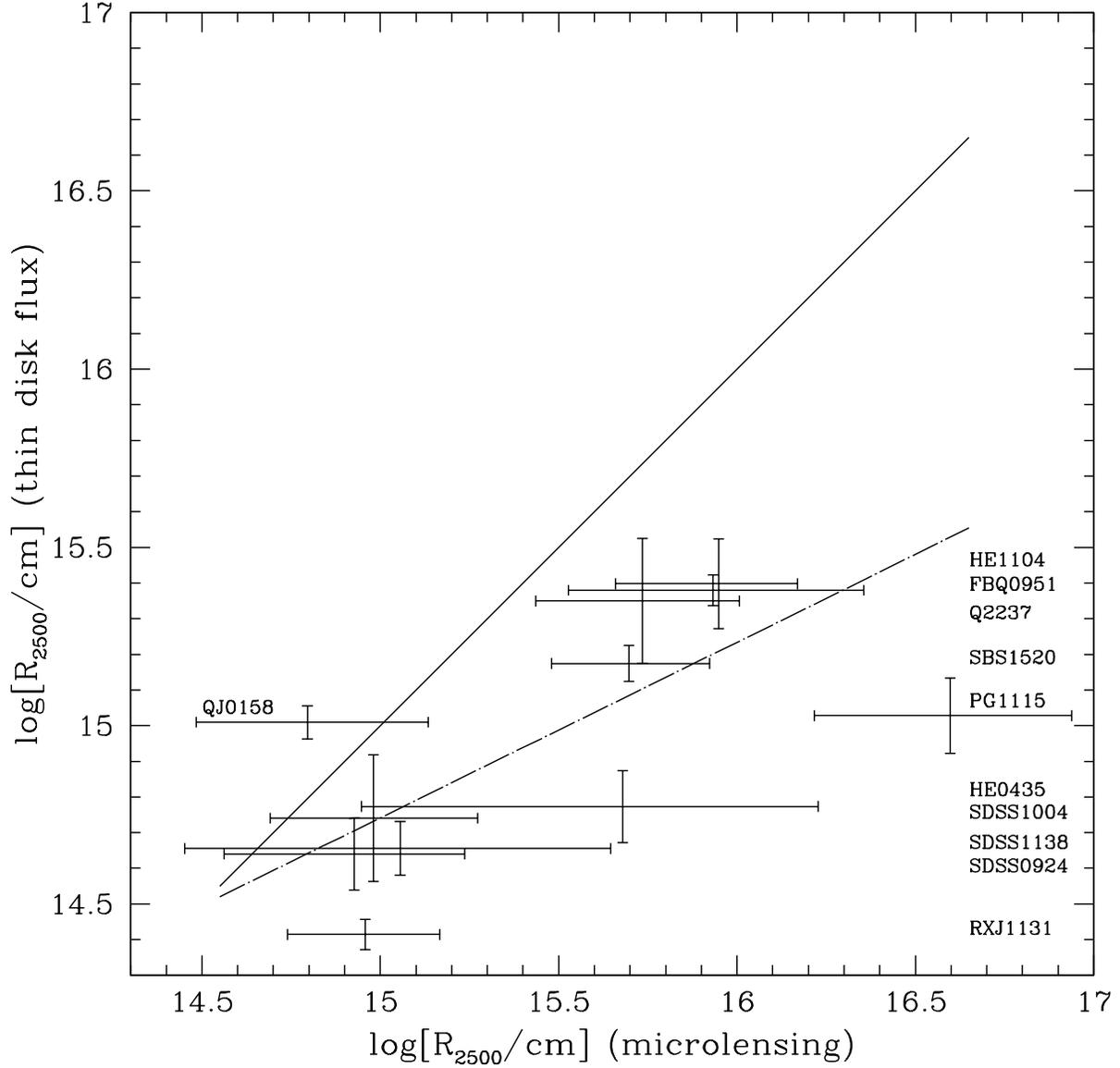}
\figcaption{Thin disk flux size estimates versus accretion disk sizes 
from microlensing.  For reference, the solid line indicates a one-to-one 
relationship between thin disk flux size estimates and the microlensing 
measurements.  The dot-dashed line is the best fit to the data. Since the data 
points have large errors relative to their dynamic range, the best-fit slope is 
consistent with unity and its average offset from the solid line is $0.6$~dex.
\label{fig:rflux}}
\end{figure}

Our results are shown in Figures~\ref{fig:r2500} through~\ref{fig:rflux} 
and summarized in Table~\ref{tab:rs}.  
For the comparison with theory and the figures, we corrected the measured sizes to 
$\lambda_{rest} = 2500{\rm \AA}$ assuming the $\lambda^{4/3}$  scaling of thin disk theory 
and a mean inclination $\langle \cos i \rangle = 1/2$. We chose 2500\AA\ because it was typical
of the actual rest-frame wavelength (see Table 1), minimizing the sensitivity of our estimates 
to any uncertainty in the true wavelength scaling.  Only the size of RXJ1131--1231 is strongly affected by 
changing the scaling of size with wavelength, because of its remarkably low source redshift ($z_s=0.658$).
However, even if we vary the temperature profile of the disk from $T \propto R^{-1/2}$ 
\citep[e.g.][]{Francis1991} to $T \propto R^{-1}$, corresponding to the range from $R_\lambda \propto \lambda^2$ to 
$R_\lambda \propto \lambda$, the wavelength-corrected disk size changes only by $\sim25\%$ and the
fit parameters for the relationship between disk size and black hole mass change by less than $5\%$.

Although we use a face-on disk model, 
we need to consider the role of inclination. Both the microlensing
size and the flux size (Eqn.~\ref{eqn:fluxsize}) are set by the projected area of the disk, so 
the true microlensing disk scale should be $1/\sqrt{\cos i}$ of the face-on estimate. When we average over
an ensemble of systems assuming a random distribution of inclination angles, we are averaging over projected areas
$\langle R_{fit}^2 \rangle = \langle R_{true}^2 \cos i \rangle = (1/2 \pm 1/2\sqrt{3})R_{true}^2$,
so we correct our measurements to $\langle R_{true}\rangle = \sqrt{2} R_{fit}$.
We discuss the consequences of other inclination distributions in \S~\ref{sec:discussion}. 
In Table~\ref{tab:rs} and Figs.~\ref{fig:r2500} and~\ref{fig:rflux}, 
we use this average correction for both the microlensing and flux sizes.  
The gray band in Fig.~\ref{fig:r2500} shows the expected extra variance of $1/2\sqrt{3}$ arising from the
inclination if we view our fits as matching the predicted and observed projected areas.
 
There are two striking facts illustrated by the figures.  First, we clearly see 
from Fig.~\ref{fig:r2500} that the microlensing sizes are well correlated with the black hole mass. 
A power-law fit between $R_{2500}$ and $M_{BH}$ including the uncertainties in both quantities yields
\begin{equation}
\log \left( {R_{2500} \over {\rm cm}} \right)= (15.78\pm0.12) + (0.80\pm0.17)\log 
\left( {M_{BH} \over 10^9{\rm M_{\sun}}} \right),
\end{equation}
which is consistent with the predicted slope from thin disk theory ($R \propto M_{BH}^{2/3}$).
If we fix the slope with mass to 2/3 (see Fig.~\ref{fig:r2500fit}), the relation
implies a typical Eddington factor of $\log(L/\eta L_E)=1.77\pm0.29$.  
\citet{Kollmeier2005} estimate that the typical quasar has $L/L_E \approx 1/3$, 
which would indicate a radiative efficiency of $\eta = L / (\dot{M} c^2) \simeq 0.006$. This efficiency is 
very low compared to standard models \citep[e.g.][]{Gammie1999} and observational constraints on radiative
efficiency derived from the local black hole mass and quasar luminosity functions \citep[e.g.][]{Yu2002,Soltan1982}.
We discuss this point further in \S~\ref{sec:discussion}.

The goodness of fit, $\chi^2 = 9.15$ for $9$ degrees of freedom ($\chi^2/N_{dof}=1.07$) 
without including any effects from the spread in inclination, suggests that the errors on the size and $M_{BH}$
estimates are appropriate. While the formal uncertainties in $M_{BH}$ from the line width relations are only $\sim0.1$~dex, the 
systematic uncertainties are generally believed to be closer to $0.33$~dex  
\citep{McLure2002,Kollmeier2005,Vestergaard2006,Peng2006}. If we fit the relation assuming that this is an
additional random error in each black hole mass, then we find 
$\log \left( {R_{2500} / {\rm cm}} \right)= (15.9\pm0.2) + (1.0\pm0.3)\log 
\left( {M_{BH} / 10^9{\rm M_{\sun}}} \right)$ with a goodness of fit $\chi^2/N_{dof}=0.59$.  This implies either that 
$0.33$~dex of random uncertainty for each individual black hole mass is too large or
that we have over-estimated the uncertainties in the size measurements.  

The low efficiency estimate is closely related to the argument by \citet{Pooley2007} that the sizes 
estimated from microlensing are larger than expected from thin disk theory, but the comparison is not
exact because \citet{Pooley2007} used black hole masses determined largely
from estimates of the bolometric luminosity rather than the emission line width method.  
This causes some covariance between 
their theory and flux sizes so an exact comparison to our results is difficult. Nonetheless,
compared to \citet{Pooley2007} we find a smaller discrepancy because our full calculations of the 
microlensing sizes tend to be somewhat smaller (by an average of 0.2~dex for our four common systems)
than those of \citet{Pooley2007}.  Despite these differences, our low estimate for the 
radiative efficiency is essentially the same as the problem pointed out by these authors.

Second, there is a much more striking discrepancy between both the
microlensing and theory size estimates (Eqn.~\ref{eqn:thindisk})
with the flux sizes (Eqn.~\ref{eqn:fluxsize}).  While most of the offset 
between the microlensing size measurements and the expectations from thin disk
theory could be explained by the existing uncertainties, the offset from the estimate based on the 
quasar flux is more significant. The measured disk sizes are significantly larger than the
flux sizes in all systems except QJ 0158--4325 (see Fig.~\ref{fig:rflux}), with an average offset of
$0.6\pm0.3$~dex. Simply put, the quasars are not 
sufficiently luminous to have the sizes estimated from microlensing while
radiating as black bodies with a $T \propto R^{-3/4}$ temperature profile. 

\section{Discussion}
\label{sec:discussion}

Using microlensing we have made the first observational demonstration of the dependence 
of accretion disk sizes on black hole mass.  While the slope of the relation is still
relatively uncertain, it is consistent with the scaling of $M_{BH}^{2/3}$ expected
for quasars with similar Eddington ratios.  The absolute scales correspond to
relatively low radiative efficiencies, which we discuss in detail below.

Like \citet{Pooley2007} we find that the microlensing estimates of
the disk size are somewhat larger than would be expected from thin
disk theory for typical Eddington ratios ($L/L_E \sim 1/3$, e.g.
\citealt{Kollmeier2005}).  We can examine this in terms of our estimate
of the radiative efficiency
\begin{equation}
  \log \eta = (-2.25 \pm 0.29) + \log \left( 3 { L \over L_E} \right) + 2 \log 
  \left( {M_{BH,true} \over M_{BH,est}} \right) + {3 \over 2} \log( 2 \, \langle \cos i \rangle),
\label{eqn:efficiency}
\end{equation}
where $3 L/L_E$ is the Eddington ratio normalized to $1/3$, $M_{BH,true}/M_{BH,est}$ is
the average ratio of the true black hole masses to our estimates,
and $2\langle \cos i \rangle$ is the dependence on changes in
the mean inclination angle from that for a uniform distribution.
Constraints derived from matching the total radiated energy of quasars
to the local black hole mass function \citep[e.g.][]{Soltan1982,Yu2002} argue for a
typical radiative efficiency close to 10\% or $\log \eta = -1$ and even higher 
efficiencies for the most luminous quasars.  It is
barely possible to reconcile our estimates with these.  First, in
unification models \citep[e.g.][]{Antonucci1993,Urry1995} the inclination
angles of optical quasars are preferentially face-on rather
than uniformly distributed.  Indeed, the first microlensing
measurements of a disk inclination angle favor face-on viewing
angles \citep{Poindexter2010b}.  If, for example, we assumed quasars
were uniformly distributed only over the range $1/2 < \cos i <1$,
then our estimate of the efficiency rises by $0.2$~dex.  The
methods used in \citet{Kollmeier2005}, or any other study of
Eddington factors, will generally be more reliable for the
distribution of the Eddington factors than for the absolute
values.  The absolute values are sensitive to the absolute
calibration of the black hole masses and correctly estimating
the bolometric luminosity.  Shifting the typical quasar from
$L/L_E=1/3$ to $L/L_E \simeq 1$ would reduce the discrepancy
by $0.5$~dex.  Similarly, most studies of estimating black hole
masses from emission line widths would accept that there are
absolute calibration uncertainties of order $0.3$~dex in
$M_{BH,true}/M_{BH,est}$.  Unfortunately, we gain only this
factor rather than its square if we make $M_{BH,true}/M_{BH,est}=3$, 
despite Eqn.~\ref{eqn:efficiency}, because the estimates of the Eddington factor also
have to be adjusted downwards if we raise the estimated black
hole mass.  If we move all three of these terms in the same
direction ($L/L_E=1$ instead of $1/3$, $M_{BH,true}/M_{BH,est}=3$
instead of $1$ and $\langle\cos i \rangle = 3/4$ instead of $1/2$)
we can we bring $\log \eta = -1.29 \pm 0.29$ within $1\sigma$ of
10\% efficiency.

\begin{figure}
\epsscale{1.0}
\plotone{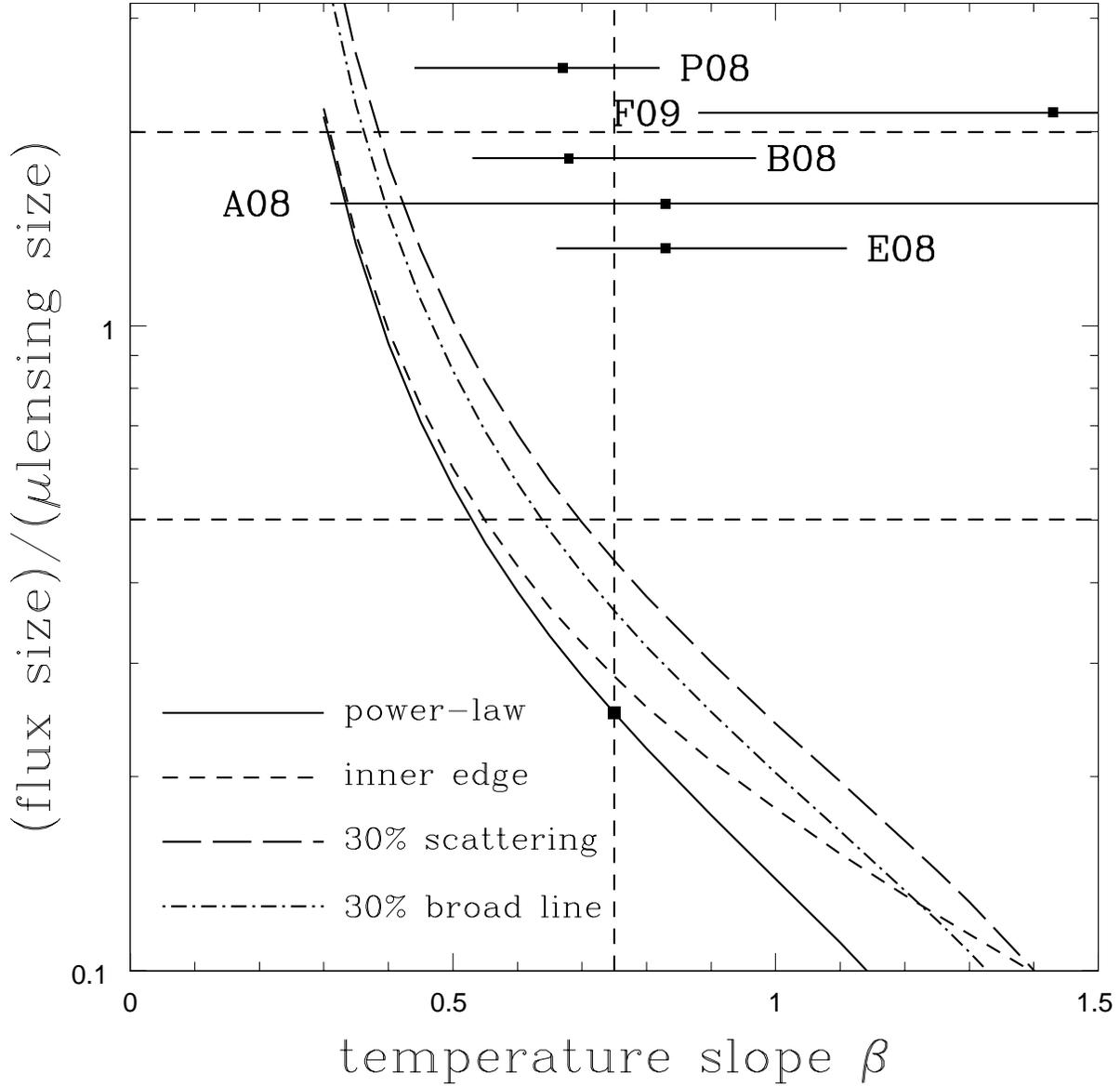}
\figcaption{Ratio between the flux and microlensing size estimates as a function
 of the temperature profile slope $\beta$, where $T\propto R^{-\beta}$.  The
 fiducial power-law model (solid) is normalized by the observed ratio at
 the $\beta=3/4$ slope of the thin disk model.  The ratio must be shifted
 into the region delineated by the horizontal dashed lines for the two sizes
 to agree given the uncertainties.  Also shown are the effects of adding
 an inner disk edge at $R_{in}=0.1R_\lambda$ (short dashed curve), making 30\% of the emission
 contamination from larger physical scales such as the broad line region (dot-dashed curve),
 and scattering 30\% of the disk flux on larger physical scales (long dashed curve).  Points
 with error bars show limits on $\beta$ from P08 \citep{Poindexter2008},
 F09 \citep{Floyd2009}, B08 \citep{Bate2008}, A08 \citep{Anguita2008} 
 and E08 \citep{Eigenbrod2008}.
\label{fig:scaling}}
\end{figure}

This may be stretching the allowed parameter shifts, but it does
not address the still larger discrepancies between either the
microlensing or the theory sizes with the flux sizes.  To
emphasize this point, if we estimate $\eta$ from the flux sizes
we find  $\log \eta = -0.43\pm0.21$ which has problems of the opposite sign.  Are these discrepancies
between these three size estimates due to a problem in the measurements,
an oversimplification of the disk model or a fundamental problem in
the thin disk model? We have tested our 
approach using Monte Carlo simulations of light curves and verified that we recover the 
input disk sizes.  Our results are also only weakly sensitive to the assumed prior on the 
microlens masses \citep[see][for a discussion]{Kochanek2004,Morgan1115}. \citet{Dai2009} in their
detailed study of microlensing in RXJ 1131$-$1231 show that none of the details of our approach
significantly affect the size estimates.  This suggests that we must
look to significant changes in the physical model to explain the differences.  Fig.~\ref{fig:scaling}
illustrates the consequences of four possible modifications: scattered light, contaminating
emission, effects of the inner disk edge, and changes in the temperature profile.  In
each case we rescale the ratio of the flux and microlensing sizes while holding the
optical flux and the half light radius of the disk fixed to see if we can shift the
ratio from the observed $-0.6\pm0.3$~dex.   Holding the half-light radius fixed
should mean that the new model will be consistent with the microlensing constraints
\citep{Mortonson2005}.  We generalize the temperature profile
of Eqn.~\ref{eqn:disk1} to 
\begin{equation}
   T(R) \propto \left( { R_\lambda \over R }\right)^\beta \left[ 1 - \left( { R_{in} \over R} \right)^{1/2}\right]^{1/4}
\label{eqn:betamodels}
\end{equation}
with locally thermal emission for $R>R_{in}$.  Our fiducial models have $\beta=3/4$
and $R_{in}=0$.  Here $R_\lambda$ is just a scale length and does not correspond
to the expression in Eqn.~\ref{eqn:thindisk}. Because Eqn.~\ref{eqn:betamodels} is not
based on a theoretical model, we have no means of relating $R_\lambda$
to physical parameters. We can attempt to reconcile the flux and microlensing sizes with Eqn.~\ref{eqn:betamodels}, 
but we cannot use it to examine the efficiency problem.

We have assumed so far that the observed optical flux is emission directly from the 
accretion disk, hence we will overestimate the microlensing size of the disk if an 
appreciable fraction of the emission originates on scales larger than the accretion
disk.  For example, if we fit the data modeling 30\% of the observed light as
unmicrolensed emission from scales much larger than the accretion disk, 
then we find that the microlensing size estimates shrink by
20-50\%.  \cite{Dai2009} investigated this problem
in detail for RXJ~1131$-$1231. As the fraction of contaminating light
increases, the disk has to become more compact in order to produce the same 
amplitude of microlensing variability.  This large scale emission can be
either scattered emission from the disk or contaminating emission from some
other source.  The two effects differ because the flux from the disk is
unchanged by scattering, while the flux from the disk is reduced by
the amount of contaminating emission.

The main sources of contamination will be emission from the larger and minimally microlensed 
line emitting regions \citep[e.g.][]{Sluse2007,Sugai2007} and the unmicrolensed emission
of the quasar host galaxy. Here we need not worry about the host galaxy as we are
examining the rest frame ultraviolet emission of luminous quasars.   
The sources of line contamination include not only the obvious 
broad/narrow lines but also the broad \ion{Fe}{2} and Balmer continuum emission 
\citep{Maoz1993,Vestergaard2005} that can represent $\sim30\%$ of the apparent 
continuum flux at some wavelengths \citep{Netzer1983,Grandi1982}.  Table~\ref{tab:rs} notes
the possible sources of broad line contamination for each system.  While four of
the systems have the \ion{Mg}{2} line in the filter band pass, the Balmer continuum and Iron
line complexes are probably the dominant source of line contamination.  However, 
the line emission is generally reprocessed harder radiation, so as we increase
the line contamination to reduce the microlensing size of the disk, we also reduce 
the optical flux coming from the disk and hence the flux size of the disk.  
Thus, producing the large scale emission by scattering the disk emission 
reduces the microlensing size but leaves the flux size unchanged, while producing
it by line contamination reduces both size estimates.  Fig.~\ref{fig:scaling} shows that 
in our simple model, contamination and scattering cannot bring the two sizes into 
agreement unless most of the emission is not coming directly from the disk.
Our discussion of scattering here assumes
it is not accompanied by significant changes in the photon energy.
If much of the observed optical radiation is due to Compton scattering
of softer photons, then the net effect would depend on the physical
scale of the scattering medium, particularly since the highest
densities of hot electrons are likely to be on similar scales to
that of the accretion disk.
One consequence of reducing the directly observed emission from the disk is
that the fractional variability of the disk emission rises in proportion, as the 
(re)emission on large (parsec) scales cannot produce the short time scales of
the intrinsic quasar variability.  This consideration probably rules out models
in which a minority of the observed flux comes directly from the disk.   

The problem is also not a consequence of the simplifications in the disk model,
namely the neglect of the inner edge and the different inner temperature 
profile of a relativistic disk \citep[e.g.][]{Page1974}.  Whether we use
the microlensing or flux scales for the disk, the optical emission is
mostly radiated far from the scale expected for the inner edges of disk
($\sim60 r_g$ rather than a few $r_g=GM_{BH}/c^2$).  \cite{Dai2009} examined this problem for
RXJ1131$-$1231 using the full relativistic \cite{Hubeny2001} models and found 
few changes from our simple standard model.  We illustrate this here by
adding an inner edge to the disk with $R_{in}=0.1 R_\lambda$, which is 
larger than we would expect from the sizes and black hole 
masses of these sources.  As shown in Fig.~\ref{fig:scaling}, this has
little effect on the size ratio unless the disk temperature profile is
very steep. 

The final possibility we consider is changing the temperature profile 
of the disk.  As shown in Fig.~\ref{fig:scaling}, using a
temperature profile flatter than $\beta=3/4$ has the strongest effect on the 
size ratio of all the changes we consider.  A temperature slope closer 
to $\beta \sim 0.5$ can bring the two scales into agreement.  With the 
addition of a scattered or contaminating component, smaller changes are 
needed, and the inner disk edge becomes less important for
the flatter profiles.  Steeper temperature profiles, on the other hand, 
worsen the problem, although in this regime the inner disk edge also becomes
important.   

Fig.~\ref{fig:scaling} also shows the existing 
microlensing limits on $\beta$ from 
\citet{Anguita2008}, 
\citet{Bate2008},
\citet{Eigenbrod2008}, 
\citet{Floyd2009}, 
and 
\citet{Poindexter2008}, 
based on 
the wavelength dependence of the microlensing. \citet{Mosquera2009} also
report results consistent with $\beta\sim 0.75$.    
With the exception of \citet{Floyd2009}, these initial studies are consistent with 
both $\beta=3/4$ and the shallower profiles that would help to resolve the differences 
in the size estimates.  The discrepant result of \citet{Floyd2009} requiring a steeper 
slope is likely a consequence of their approach.
Unlike the other studies, both \citet{Bate2008} and \citet{Floyd2009}  
use only the color differences between images observed at single epochs rather
than light curves.  This means that at any wavelength they can only set
an upper bound on the source size -- it is the amplitude observed
during caustic crossings that sets the lower bounds, and that can only
be measured using time variability.  However, given only a set of upper limits
on the sizes that are smaller for shorter wavelengths, the estimate for
$\beta$ will be determined by the priors used for the source sizes. In
particular, the linear priors used by \citet{Bate2008} and \citet{Floyd2009}
will favor steep temperature profiles, while a logarithmic prior
would favor shallow temperature profiles.  The same problem occurs
for sparsely sampled light curves, as illustrated by the studies of
X-ray microlensing by \citet{Morgan1115} and \citet{Dai2009}, where the lower
limits to the X-ray size are found to be prior-dependent.  From this
point of view \citet{Bate2008} and \citet{Floyd2009} are really upper
bounds on $\beta$, and analyses of light curves will be required
to determine a lower bound.    

Arguments for a flatter emissivity profile in accretion disks have
existed for a long time, largely based on the mismatch between the
predicted and observed spectra of quasars (see the reviews by
\citet{Koratkar1999} and \citet{Blaes2004}).  The emission profile
can be flattened either by raising the temperature of the outer
disk (e.g. irradiation of the outer disk by the inner) or by
changing the balance between radiation and advection in the
inner disk (e.g. thick or slim disks).  Microlensing provides the first
probe able to actually measure the physical scales associated with
the emission regions, and the results are clearly beginning to 
test the simplest theories.  As the numbers of systems, the
uncertainties in the size measurements and the wavelength range
spanned by the measurements increases, microlensing can quantitatively
address all these issues.

\acknowledgments
We thank O. Blaes, E. Agol, M. Dietrich, C. Onken, B. Peterson, M. Pinsonneault, R. Pogge and 
P. Osmer for discussions on quasar structure and M. Mortonson, S. Poindexter, S. Rappaport, and P. 
Schechter for discussions on microlensing. We would also like to thank the anonymous referee for 
valuable suggestions, particularly in the 
expansion of our discussion of radiative efficiencies. This research made extensive use
of a Beowulf computer cluster obtained through the Cluster Ohio
program of the Ohio Supercomputer Center. Support for program HST-GO-9744 was
provided by NASA through a grant from the Space Telescope Science Institute, which 
is operated by the Association of Universities for Research in Astronomy, Inc., under
NASA contract NAS-5-26666.  This material is based upon work supported by the National Science Foundation under 
grant No. AST 0907848. This research was also supported 
by an award from the Research Corporation for Science Advancement. 

{\it Facilities:} \facility{CTIO:2MASS (ANDICAM)}, \facility{Hiltner (RETROCAM)}, \facility{WIYN (WTTM), 
\facility{HST (NICMOS, ACS)}}.

\clearpage

\begin{deluxetable}{lcccccccc}
\tablewidth{0pt}
\rotate
\tabletypesize{\scriptsize}
\tablecaption{Measured and Derived Quantities}
\tablehead{\colhead{Object} 
		& \colhead{Line}
		& \colhead{FWHM}
		& \colhead{$M_{BH}$} 
		& \colhead{$\log(R_S/{\rm cm})$}
		& \colhead{$\lambda_{rest}$}
                & \colhead{BLR}   
		& \colhead{$I_{corr}$}
		& \colhead{$\log(R_S/{\rm cm})$} \\
		\colhead{}
		& \colhead{}
		& \colhead{\AA~(observed)}
		& \colhead{$(10^9~{\rm M_\sun})$}
		& \colhead{(microlensing)}
		& \colhead{(\micron)}
                & \colhead{Contaminant}
		& \colhead{(mag)}
		& \colhead{(thin disk flux)}
		}		
\startdata
QJ0158--4325  & Mg~{\sc ii} & 40 & 0.16 & $14.9_{-0.3}^{+0.3}$ & 0.306 & Balmer, Fe~{\sc ii} UV, Mg~{\sc ii} & $19.09\pm0.12$ & $15.2\pm 0.1$\\ 
HE0435--1223  & C~{\sc iv} & 70 & 0.50 & $15.7_{-0.7}^{+0.5}$ & 0.260 & Balmer, Fe~{\sc ii} UV & $20.76\pm 0.25$ & $14.9\pm0.1$ \\
SDSS0924+0219  & Mg~{\sc ii} & 61 & 0.11 & $15.0_{-0.4}^{+0.3}$ & 0.277 & Balmer, Fe~{\sc ii} UV, Mg~{\sc ii} & $21.24\pm 0.25$ & $14.8\pm0.1$  \\
FBQ0951+2635  & Mg~{\sc ii} & 70 & 0.89 & $16.1_{-0.4}^{+0.4}$ & 0.313 & Balmer, Fe~{\sc ii} UV, Mg~{\sc ii} & $17.16\pm 0.11$ & $15.6\pm0.1$ \\
SDSS1004+4112  & Mg~{\sc ii}  & 134 & 0.39 & $14.9_{-0.3}^{+0.3}$ & 0.228 & Balmer, Fe~{\sc ii} UV & $20.97\pm 0.44$ & $14.9\pm0.2$ \\ 
HE1104-1805  & C~{\sc iv} & 103 & 2.37 & $15.9_{-0.3}^{+0.2}$ & 0.211 & Balmer, Fe~{\sc ii} UV & $18.17\pm 0.31$ & $15.4\pm0.1$ \\
PG1115+080  &Mg~{\sc ii}  & 127 & 1.23 & $16.6_{-0.4}^{+0.3}$ & 0.257 & Balmer, Fe~{\sc ii} UV & $19.52\pm 0.27$ & $15.1\pm0.1$ \\
RXJ1131--1231  & H$\beta$ & 90 & 0.06 & $15.3_{-0.2}^{+0.2}$ & 0.422 & Balmer, Fe~{\sc ii} Optical & $20.73\pm 0.11$ & $14.8\pm0.1$ \\
SDSS1138+0314  &  C~{\sc iv} & 25 & 0.04 & $14.9_{-0.6}^{+0.6}$ & 0.203 & Balmer, Fe~{\sc ii} UV & $21.97\pm 0.19$ & $14.6\pm0.1$ \\
SBS1520+530  &  C~{\sc iv} & 75 & 0.88 & $15.7_{-0.2}^{+0.2}$ & 0.245 & Balmer, Fe~{\sc ii} UV & $18.92\pm 0.13$ & $15.3\pm0.1$ \\
Q2237+030  &  C~{\sc iv} & 48 & ~0.9\tablenotemark{a} & $15.6_{-0.3}^{+0.3}$ & 0.208 & Balmer & $17.90\pm 0.44$ & $15.5\pm0.2$ \\
\enddata
\tablecomments{$R_S$ from microlensing is the accretion disk size at
$\lambda_{rest}$, the rest-frame wavelength corresponding to the 
center of the monitoring filter used for that quasar's light curve.
Use half-light radii ($R_{1/2}=2.44 R_S$) to compare these size measurements to other disk models.
Significant sources of unmicrolensed flux from the QSO Broad Line Region (BLR) that fall into or overlap with the
observing pass band are indicated: Balmer Continuum ($\lambda \lesssim 3650$\AA), Fe~{\sc ii} UV Continuum ($\lambda \lesssim 3100$\AA),
Fe~{\sc ii} Optical Continuum ($4240$\AA~$\lesssim \lambda \lesssim5400$\AA) or Mg~{\sc ii}~($\lambda 2798$\AA).  
$I_{corr}$ is the corrected (unmagnified) $I$-band magnitude.
Typical $I$-band measurement errors are $\lesssim 0.1$ mag, but the larger errors
on $I_{corr}$ come from uncertainties in the lens magnification.
$R_S$ calculated using corrected $I$-band magnitude and thin disk theory is also unscaled; it is the 
disk size at the rest-frame wavelength corresponding to the center of the 
{\it HST} $I$-band filter (F814W).  Both disk sizes assume an average inclination angle $i=60\degr$.}
\label{tab:rs}
\tablenotetext{a}{The  C~{\sc iv} emission line width from \citet{Yee1991} depends strongly on the fit to 
several blended  C~{\sc iv} absorption features, so we report $M_{BH}$ at lower precision.} 
\end{deluxetable}

\clearpage

\end{document}